# Systemic Risk Clustering of China Internet Financial Based on t-SNE Machine Learning Algorithm

Mi Chuanmin1, Xu Runjie1, Lin Qingtong2
1.College of Economics and Management,
Nanjing University of Aeronautics and Astronautics,
Nanjing , China
2. College of Management, Da-Yeh University, Taiwan

**Abstract:** With the rapid development of Internet finance, a large number of studies have shown that Internet financial platforms have different financial systemic risk characteristics when they are subject to macroeconomic shocks or fragile internal crisis. From the perspective of regional development of Internet finance, this paper uses t-SNE machine learning algorithm to obtain data mining of China's Internet finance development index involving 31 provinces and 335 cities and regions. The conclusion of the peak and thick tail characteristics, then proposed three classification risks of Internet financial systemic risk, providing more regionally targeted recommendations for the systematic risk of Internet finance.

**Introduction**

Internet finance makes full use of the advantages of the Internet, improves the efficiency of financial resource allocation to a certain extent, and promotes the inclusive development of finance.However, fundamentally speaking, the essence of Internet finance is still finance, which contains the hidden, infectious and sudden characteristics of financial risks and brings new systemic risks.The existing research on financial systemic risk, on the one hand, USES the data of commercial Banks or financial markets to establish the measurement model of systemic risk.For example, lling and Liu build a financial stress index (FSI) to study financial system risks.Chen shoudong and wang yan introduced the extremum theory into the measurement of systemic financial risk, thus proving the non-normal distribution characteristics of market value of financial institutions.Klafft analyzed the P2P industry and concluded that the transaction risk of Internet finance was due to the lack of experience in loan business.Onay and Ozsoz analyzed the loan approval business of Internet finance and commercial Banks, and found that Internet finance can enjoy relatively low loan interest rate. Internet finance has squeezed the profits of commercial Banks, and commercial Banks have to be stimulated to make changes in operational efficiency and profitability.Lanxiang measured the risk of Internet finance by selecting csec 800 financial index and based on VaR analysis and Copula. The results showed that the VaR and ES of Internet financial market were higher than those of traditional market, and Internet financial market itself may have higher systemic risk.Wang liyong and shi ying have adopted the two-level CRITIC- grey relation model to construct the Internet financial risk assessment system, and have used the VaR method to measure the risk of Internet finance.In addition, there are also studies on the systematic risks of Internet finance from the perspective of risk contagion.Based on the analytic hierarchy process (ahp), jia nan analyzed the influence weight of Internet finance on technical risk, operational risk, legal risk, credit risk and business risk, and concluded that credit and technology are the risk contagion factors with the highest importance in Internet finance.Through the analysis of Internet media, social media and other attributes, yao guozhang believes that Internet finance promotes its own business in the form of Internet products, leading to the spread of risks that are difficult to control.Sui cong, chi guotai, deng chao, Chen xuejun et al. used the Watts level linkage mechanics theory of complex network for reference to construct a financial contagion model based on random network.Mi chuanmin et al. constructed the super-network model of Internet finance and studied the financial system equilibrium and systemic risk contagion in the relationship between social networks in Internet finance.According to the interaction between Internet finance and banking industry, zhu Chen and hua guihong describe the trigger and contagion mechanism of systematic risk in banking industry under the influence of Internet finance, which shows that Internet finance not only interacts with commercial Banks, but also opens the contagion channel of systematic risk.

The above studies show that Internet finance not only injects new vitality into the economy, but also accompanies the process of risk accumulation.However, there are relatively few studies on the regional characteristics of Internet financial risks, including the influence of different regions on the contagion mechanism of Internet financial risks.Therefore, this paper USES the "Internet finance development index" of Peking University to empirically study the influence of different regions and business models in the process of Internet finance systemic risk induction from the perspective of risk contagion.It mainly solves the dimensional crowding problem caused by the high complexity and many variables of the Internet financial data, and adopts the method of spatial mapping to describe the data that is difficult to observe the spatial structure into a two-dimensional image that can retain local information.In financial analysis our country Internet payment, money funds, insurance, investment, four types of business, the Internet has the characteristics of geographical spatial clustering of financial development model, put forward different areas provided by different levels of risk, and Internet financial under different business model shows the characteristics of different stages of development and risk.

**The data source**

"Internet finance development index" was compiled by Peking University and major domestic Internet finance enterprises based on the development of Internet finance from January 2014 to December 2015.This paper selected 35,136 data from 31 provinces (excluding Hong Kong, Macao and Taiwan, the same below) and 335 prefecture-level cities for empirical research.This data includes different business model indicators and is reflected in each city by time stamp, with many variables and high dimensions.The traditional multivariate statistical method will encounter the situation that the data does not conform to the normal distribution when dealing with the actual data, and it is difficult to directly observe the spatial structure.Therefore, this paper adopts machine learning dimensional reduction algorithm to train a large number of data, so as to visualize the data and study the relationship between the spatial distribution of Internet finance and business development.

According to the Internet finance business, the index divides the development of domestic Internet finance into four main businesses: Internet payment, Internet fund, Internet credit and Internet insurance. The specific business indicators are determined by the weight of transaction penetration rate, transaction amount per capita and number of transactions per capita, which are respectively 50%, 25% and 25%.The index has the advantages of representativeness, operability, independence and extensibility, and is one of the few Internet finance research data in China.In the data set of this study, the Internet finance index by region is included. In the calculation process, the horizontal comparability between the index points of different regions is taken into account. The calculation formula of the three indicators of relative transaction penetration rate, transaction amount per capita and transaction number per capita in different regions in a certain period is as follows:

$$A_{h,i,j,t} = \frac{X_{h,i,j,t}}{X_{i,j,t}}$$

$A_{h,i,j,t}$ represents the relative value of the jth index of category i business in region h relative to the national total index at time t, $X_{i,j,t}$ Represents the same operation and index of the national total index at time t, $X_{h,i,j,t}$ Represents the same operation and index of region h at time t.

The coefficient calculation formula of Internet payment, fund, credit and insurance business in different regions relative to the whole country in a certain period is as follows:

$$B_{h,i,t} = \sum_{j=1}^{3} m_j A_{h,i,j,t}$$

（2）

$B_{h,i,t}$ represents the relative coefficient of category i business in region h relative to the national total index in period t. M1, m2 and m3 respectively represent the above transaction penetration rate, per capita transaction volume and the weight of the number of transactions per capita. M1 =50%, m2=25%, m3=25% in this index.

From the data set of this study, we selected the index data of 31 provincial (excluding Hong Kong, Macao and Taiwan, the same below) and 335 prefecture-level cities from January 2014 to December 2015 in the four major businesses of Internet payment, cargo base, credit and insurance respectively as the empirical data of this topic.

**t-SNE algorithm**

t - distributed stochastic neighbor embedding （ t - SNE), by the Laurens van der adopted Maaten and Geoffrey Hinton, the late after the adopted Maaten improvement, respectively in 2015 and 2016 LINE and LargeVis algorithm is proposed, and greatly reduce the training complexity.At present, this algorithm has achieved good results in the application of dimensionality reduction, clustering and visualization. For example, Gordon Berman applied t-SNE algorithm to the dimensionality reduction analysis of the video of the free movement of fruit flies on the ground (that is, in addition to flight).Feng rui and yuan ruiqiang introduced dimension reduction technology into the field of water quality evaluation, and discussed the water quality evaluation method based on t-SNE algorithm.Liu feng et al. realized the data visualization system of bird audio intelligence identification based on t-SNE algorithm.Zhan weiwei and wang bin et al. used t-SNE algorithm in the state observation matrix of high-dimensional brain network, which effectively solved the dispersion, intersection and scatter.The provincial and prefecture-level city data selected in this paper are similar to the research data in the above literatures, both of which have high data dimension and complex nature, including four different business models, more than 300 measurement regions and 24 time nodes.In many classic data processing methods, such as regression analysis, principal component analysis and correlation analysis, for sparse and multi-variable complex data sets, they can only reflect the overall correlation and ignore the local connection.However, t-SNE algorithm USES the joint probability of high and low dimensions to find the attraction between points, which can effectively solve the problem of optimization difficulty and dimensional crowding.In this way, the Internet finance index can be mapped into a two-dimensional image highlighting the regional distribution. The reserved spatial structure will be more conducive to the observation of the development of the entire Internet finance and the study of its risks.

t-SNE algorithm mainly focuses on the local properties of data, and realizes the optimization extraction of data by using the method of combining high dimension with low dimension, so as to present the complex nonlinear relationship of local data.In high-latitude space, t-SNE algorithm adopts gaussian distribution and uses probability $p_{i|j}$ to express the relationship between I point and j point:

$$P_{i|j} = \frac{\left(1+\|y_i-y_j\|^2\right)^{-1}}{\sum_{k \neq l}\left(1+\|y_k-y_l\|^2\right)^{-1}}, \quad \forall i,j \text{ 且 } i \neq j$$

In low dimensional space, t distribution with 1 degree of freedom is adopted，which is represented by $q_{i|j}$.

$$q_{i|j} = \frac{q_{i|j} + q_{j|i}}{2n}, \quad \forall i,j \text{ 且 } i \neq j$$

In the process of dimensionality reduction extraction of complex data, the optimal state is that thesimilarity between high-latitude spatial sample points is the same as that between low-dimensional spatial sample points. t-SNE uses kullback-leibler divergence as the objective function to judge the difference, so as to achieve the optimal parameter.

$$C = KL(P\|Q) = \sum_i \sum_j p_{j|i} \log \frac{p_{i|j}}{q_{i|j}}$$

The minimization iterative formula achieved by gradient descent method is as follows:

$$\frac{\delta C}{\delta y_i} = 4 \sum_j (p_{i|j} - q_{i|j})(y_i - y_j)\left(1 + \|y_i - y_j\|^2\right)^{-1}$$

In the process of experiment with four factors affecting the effect of dimension reduction, respectively is perplexity, early exaggeration factor, vector (learning rate),maximum number of iterations(maximum number of iterations). Among the four factors mentioned above, perplexity is the most critical factor influencing the later results. Its function is to obtain the variance of the gaussian distribution. The degree of confusion of any row of the conditional probability matrix P can be defined as：

$$Perp(P_i) = 2^{H(P_i)}$$

$H(P_i)$ is the shannon entropy of $P_i$ , If the entropy of probability distribution in high-dimensional space increases, the variable uncertainty of data set will increase, thus resulting in more flat distribution of processed data. On the contrary, the lower the entropy of probability distribution in high-dimensional space, the higher the correlation of data and the more distributed the data obtained after processing. The probability distribution relationship between shannon entropy and high-dimensional space is as follows:

$$H(P_i) = -\sum_j p_{i|j} \log_2 p_{i|j}$$

**Provincial risk analysis**

In order to test the ability of the t-SNE algorithm used in this paper in processing multidimensional and multivariable complex data, the classical PCA (Principal Component Analysis) was also used in this paper for comparative study. Dimensionality reduction, clustering, imaging and Analysis were conducted on the values of 31 provincial regions (excluding Hong Kong, Macao and Taiwan) to find out a more suitable research method.

As a classical data mining algorithm, PCA is applied in many data mining scenarios. Its principle is to simplify data through linear projection, so as to map high-dimensional data to low-dimensional space and retain most internal information of the original data as much as possible.In this paper, PCA is used to reduce the multidimensional business data of each province to 2 dimensions and visualize it on the coordinate axis.The obtained imaging effect is shown in figure 1 (a). The abscissa represents the development volume index of each province, and the larger the value is, the more advanced the development will be.The vertical axis represents the degree of equilibrium turbulence in the development of the four major businesses of Internet payment, insurance, cargo base and credit. The larger the value is, the more unbalanced the development volume of the four major businesses will be, that is, the more diverse they will be.

According to the PCA analysis results, the data of the 31 provincial-level regions show the following characteristics: each province develops rapidly and most provinces gather together.However, from the perspective of the equilibrium degree of different business development of Internet finance, the regions with excellent overall development of Internet finance also have low differences in business development, which may be due to the perfect Internet financial infrastructure, high Internet penetration rate, strong economic vitality and consumption level in developed provinces.According to the figure, the regions with low overall development degree of Internet finance also show the balanced development between different businesses, which may be caused by the lack of Internet economic development momentum in these regions.Compared with the former two, there is a phenomenon of unbalanced development among businesses in cities in the middle of development, which precisely illustrates the great potential of Internet finance in developing cities and makes the tail dependence particularly important in the accumulation of systemic risk.

According to the numerical results processed by PCA, sort them by size. Some of the fastest growing cities, such as Beijing, are eight times higher than the national average.In addition, with the development speed of Internet finance far exceeding that of inland cities in the eastern coastal provinces, most of the western provinces of China are still at the end of the development of Internet finance.The ranking also reflects the need to consider the weight of individual cities in the field of Internet finance in the consideration of the systemic risk of Internet finance.

This paper USES python to implement t-SNE algorithm to analyze risks of Internet financial system in 31 provincial regions.Since t-SNE projects high-dimensional spatial samples of data onto multiple two-dimensional spatial maps starting from the non-linear dimensionality reduction method, and USES the bidirectional probability distribution of high and low dimensions, the values obtained by t-SNE algorithm have the characteristics of randomness.In this paper, the popular regularization technique is combined to control the map, so that the sample points projected into the visualized space can not only maintain the overall structure of the high-dimensional data, but also maintain the relationship between local neighbor points.In the t-SNE data processing experiment, it is mainly divided into two stages, namely the preliminary amplification stage and the performance selection stage.In the experiment, the preliminary amplification coefficient was set as 12.0, and the joint probability of the space was gradually increased by multiplying the preliminary amplification coefficient.The key point is to adjust the degree of confusion and the learning rate.The distribution morphology and clustering effect under different parameters are shown in table 1 below.

**Table 1.** *Distribution under different parameters*

| Algorithm | t-SNE | | | | | |
|---|---|---|---|---|---|---|
| perplexity | P=20.0 | | P=10.0 | | P=5.0 | |
| Number of training | L=200 | L=500 | L=200 | L=500 | L=200 | L=500 |
| Distribution form | Uniform distribution | Uniform distribution | Clustering distribution | Discrete distribution | Clustering distribution | Clustering distribution |

The clustering distribution of 31 provinces is more stable when P=5, and the clustering effect is shown in figure 1 (b) and (c). The image shows the four-level clustering division of 31 provincial regions. In this paper, based on the number of data points included in the clustering image, it can be roughly divided into:

I level: Beijing, Shanghai, Zhejiang and Guangdong.
II level: Jiangsu, Hubei, Shandong, Fujian, Tianjin, Chongqing, Liaoning.
III level: Shanxi, Hainan, Anhui, Sichuan, Shanxi, Hebei, Jiangxi, Heilongjiang, Henan.
IV level: Jilin, Xinjiang, Ningxia, Hunan, Tibet, Guangxi, Inner Mongolia, Yunnan, Qinghai, Guizhou, Gansu.

Compared with the image observation of PCA calculation results in the previous section, the image visualization based on t-SNE results is also more detailed and persuasive. In the clustering results, the provinces with strong development, the provinces with strong development, the provinces in the middle of development and the provinces with poor development are included. The clustering effect reflects the development characteristics of China's domestic Internet finance: extreme development of individual provinces and sharp peak and thick tail of most provinces gathering together.

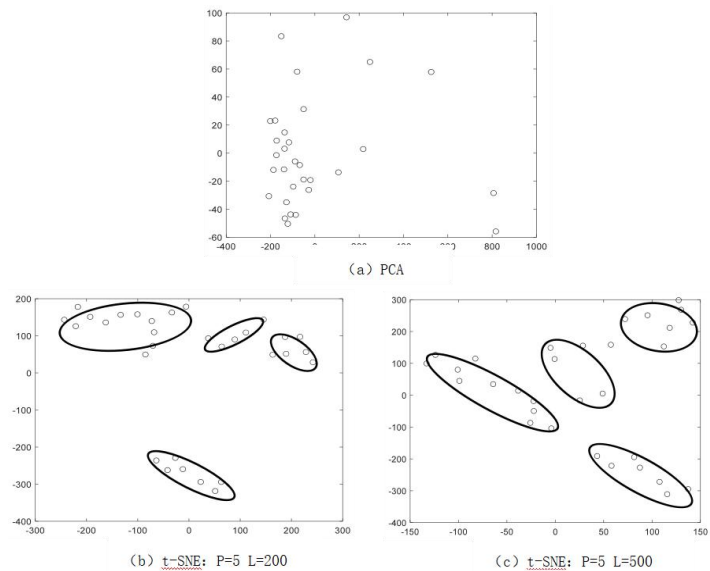

(a) PCA

(b) t-SNE: P=5 L=200

(c) t-SNE: P=5 L=500

**Figure 1.** The clustering effect of the two algorithms is compared

According to figure (1), PCA algorithm can clearly reflect the level of Internet finance in each province, so as to explain the proportion of different regions in inducing the systematic risk of Internet finance. In addition, PCA accelerates the calculation of paired distance between data points and inhibits the singular value.However, after the imaging of PCA, the development values of Internet finance in different cities show a disordered mixture of subsets, which cannot well represent the statistical characteristics required in this paper.Some cities are attracted to each other, while other cities are separated from each other at a distance, so it is difficult to intuitively observe the relationship characteristics of different regions in the development degree of Internet finance.

t-SNE dimension-reduced clustering results present clear clustering clusters, thus making the definition between the development degrees of different Internet financial regions more obvious.As a non-linear algorithm, t-SNE plays a role in noise reduction by restoring the popular structure of data in the low-dimensional state, and embodies the internal relationship of data, so as to better reflect the system characteristics before dimensionality reduction.The successful clustering of 31 provinces by t-SNE reflects the interregional correlation in the development process of China's Internet finance, which provides a data-driven reference for regulators to grade and subregional supervision of the systematic risk of Internet finance.

**The city of analysis**

In the third part, we mainly analyze the data from the level of provincial data.However, the cross-section of provincial data is only 31 categories, which cannot reflect the characteristics of Internet finance systemic risk in more details.This section will use the multi-service development data of 335 prefecture-level cities (including autonomous prefecture, league and region) for experiment and analysis.In the third part, PCA analysis is applied to avoid dimensional crowding and data mixing in the process of data processing, and it is difficult to generate clustering effect. Therefore, this section mainly discusses the system risk clustering effect of the development of Internet finance at the prefecture-level and city-level based on t-SNE machine learning algorithm.

In the t-SNE machine learning process of prefecture-level cities, the parameters are set as follows: the preliminary amplification coefficient is 12.0, the confusion degree is 30.0, and the training times are 5000. Then k-means algorithm is used to cluster the data after dimension reduction, and the clustering results are shown in figure 2.It can be observed from figure 2 that the development of Internet finance in prefecture-level cities gathers into 7 categories, and the points of one prefecture-level city are scattered near the 7 major categories.For the points of prefectural cities with large similarity, the distance of t distribution in the low-dimensional space is a little smaller, that is, the points within the same cluster (the distance is relatively close) converge more closely.For the points of prefecture-level cities with low similarity, the distance of t distribution in the low-dimensional space needs to be further, that is, the points between different clusters (the distance is relatively far) are more distant.

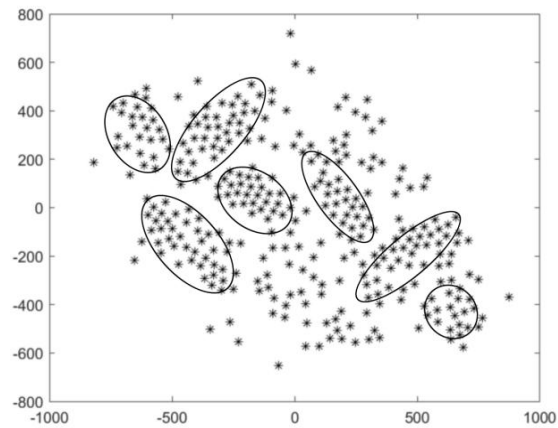

**Figure 2.** The clustering effect of the two algorithms is compared

In the process of using t-SNE machine learning algorithm to conduct experiments on sample data, through parameter adjustment, we found that: when the confusion degree of probability distribution P=10, the overall hybridity of data distribution is generated, as shown in figure 4 below. This shows that, first, the aggregation distribution of Internet finance development among regions is not complete. Although it mainly follows a whole, there are also scattered urban and spatial clustering differences. Second, it is observed in figure 3 that the regional distribution of Internet finance development in prefecture-level cities spreads outward from the origin of horizontal and vertical coordinates. Abscissa represents the relative to the size of the average level of the cities the development of Internet, the ordinate said Internet financial payment, funds, credit, insurance, the balance of the four points of business development degree of turmoil, this has to do with the PCA algorithm to get the similar results, also with the Internet financial development index local Moran's l scatter diagram to get the similar results. Most of the data floating outside the focusing origin are distributed in the first, third and fourth quadrants, and a few in the second quadrant. The specific meaning of the coordinate of the third image is: the development index of Internet finance is low, and the difference of business development is high, but the experimental result is that the second quadrant presents very little scatter distribution. Combined with the actual situation of China's Internet finance development and the above PCA clustering results, it can be seen that the balance of business development in western Chinese cities is the same as that in developed cities when the overall level of Internet finance is low. This also confirms that the development situation of China's Internet finance obtained in the third part of this paper is the extreme development of individual provinces and the sharp peak and thick tail phenomenon of most provinces gathering together.

**Management implications**

On the basis of the above research results, combined with li jianping's research results on risk diversification effect and integration, from the perspective of management enlightenment, this paper believes that the prevention and control of systemic risk in the development of China's Internet finance can be focused on from the following three aspects:

(1) risks of geographic spatial clustering infection

In order to prevent the systemic risk of Internet finance in China, we should pay attention to regional prevention and avoid regional contagion risk.From the above clustering results, it can be seen that the development of China's Internet finance is mainly concentrated in the eastern coastal areas, showing regional characteristics.On the one hand, it increases the infectivity of systematic risks in Internet finance; on the other hand, it also reflects the necessity of targeted supervision on the gathering areas.To be specific, the developed eastern regions show the extreme characteristics of the Internet development speed. Such regions have a high level of economic and financial development, complete Internet infrastructure construction and sufficient demand and supply for the development of Internet finance.All of these have greatly promoted the inclination of the development of Internet finance in spatial and regional selection.Regions with good economic development have a high growth rate in the development of Internet finance, which is accompanied by the development differences between different businesses of Internet finance.Although the economically underdeveloped western region is affected by the advantages of the Internet, which is convenient and fast across time and space, the speed and power of the development of Internet finance lags behind that of the eastern region.This regional difference in the development of Internet finance may bring about the contagion risk of Internet finance, which needs to be paid special attention to in the process of macro regulation.

(2) differentiated risks in business development

Differences in business development may cause systemic risks at the intermediate level, which needs attention.From the perspective of the development of graded businesses, the cities with the leading development of Internet finance have balanced development in the four major businesses of Internet payment, fund, credit and insurance, while the cities with the medium level of development show the characteristics of uneven development among businesses.This is consistent with Allen's research, that is, to explain systemic risk from the perspective of economic balance, considering that economic development is not only reflected in the gap in the overall volume, but also needs to refer to the development situation of markets at all levels.According to our research, from the perspective of the internal connection of the system, the unbalanced business development of cities in the middle of the development of Internet finance means that there are market differences in the development of Internet finance, which may also bring about systematic risks. Therefore, it is necessary to pay attention to the differences in business development in the process of supervision.

(3) systemic risks in the development of Internet finance in major cities

From the perspective of network science, there are connections between nodes in the network. Different nodes have different importance and influence on the network.If the national Internet finance is regarded as a network, according to the above clustering empirical study, it is found that Beijing, Shanghai, shenzhen, hangzhou and other cities are the cities with prominent development of Internet finance in China, and the financial radiation effect on surrounding areas is also the strongest.Beijing and Shanghai are based on traditional financial centers, shenzhen is based on

WeChat-related Internet finance, and hangzhou is based on ant finance, providing good opportunities for the development of Internet finance in these cities.Zhu xiaoqian et al. found that the crisis of a single financial institution may lead to the crisis of the entire financial system, and the probability can be used to measure the systemic risk.Basel III proposed the concept of systemically important financial institutions in response to the 2007 financial crisis and strengthened the supervision of systemically important institutions.On the basis of part 4 research results, we think, considering the Internet across time and space, the agglomeration effect is stronger and the characteristics of risk spread faster, in addition to focus on systemically important financial institutions, also need to focus on systemic important Internet financial development of the city, therefore, puts forward the concept of the Internet financial development systemically important cities.It is necessary to focus on the systemic risks that systemically important cities may bring to the financial system from the medium level.

**Conclusion**

Internet finance has brought profound influence to the financial field, and its risk cannot be underestimated.Different from the traditional research method of systematic grace and risk, this paper conducted reduction and clustering processing on high-dimensional data from the perspective of risk contagion, so as to study the spatial and geographical visualization of Internet finance and the structural differences in business distribution, and obtained the regional characteristics of systematic risk of Internet finance development.From the empirical results, t-SNE algorithm can better capture the characteristics and weak links of the occurrence of systemic risk, and conclude that the Internet financial market has the characteristics of peak and thick tail, on this basis, this paper proposes three types of Suggestions for the prevention and control of Internet financial systemic risk.